**Title**

Field canalization using anisotropic 2D plasmonics

**Authors**

Po-Han Chang,[1]* Charles Lin[1], Amr S. Helmy[1]*

**Affiliations**

[1]Electrical and Computer Engineering, University of Toronto, Toronto, Canada
*pohan.chang@mail.utoronto.ca

**Abstract**

Optical devices capable of suppressing diffraction nature of light are of great technological importance to many nanophotonic applications. One important technique to achieve diffractionless optics is to exploit field canalization effect. However, current technological platforms based on metamaterial structures typically suffer from strict loss-confinement tradeoff, or lack dynamic reconfigurability over device operations. Here we report an integrated canalization platform that can alleviate this performance tradeoff. It is found that by leveraging material absorption of anisotropic 2D materials, the dispersion of this class of materials can flatten without increasing propagation losses and compromising confinement. The realization of such plasmon canalization can be considered using black phosphorus (BP), where topological transition from elliptic to hyperbolic curves can be induced by dynamically leveraging material absorption of BP. At the transition point, BP film can support long range, deeply subwavelength, near-diffractionless field propagation, exhibiting diffraction angle of 5.5°, propagation distance of $10\lambda_{spp}$, and $\lambda_{spp} < \lambda_0/300$.

**Introduction**

Technological platforms that can suppress the diffraction of light have been studied extensively as they can be utilized in applications that are restricted by diffraction limit such as near-field spectroscopy, microscopy, energy harvesting, construction of self-trapping devices and free space optical communications [1-5]. For these applications, the diffraction angle along with propagation loss incurred to alleviate diffraction, are the two key attributes that dictate the performance of such nearly-diffractionless optical devices. To date, implementations of low-loss, highly collimated beams typically rely on one of two mechanisms: nonlinear self-trapping [6], and the propagation of non-diffracting Bessel beams [7, 8]. For nonlinear self-trapping, the expansion of optical beams is delicately balanced by nonlinear-induced self-focusing, which collectively permit the beam to propagate without spatial diffraction. Nonetheless, the operation of such spatial soliton requires high input optical powers and is only limited to specific materials endowed with optical nonlinearity. On the other hand, non-diffracting Bessel beams can propagate with no need for nonlinear materials. However, the excitation of Bessel beams requires a delicate control over the phase and amplitude of the wave-fronts forming the beams [9]. This inevitably increases the complexity of the system and its operation, particularly in guided wave settings that are compatible with integrated photonics. Extra structures such as subwavelength nanostructured elements [10] or photonic crystal structures [11, 12] need be deployed to assist in the generation of such beams. These resonant structures inherently impact device operational bandwidth and do not offer the ability to dynamically tune the optical properties, making them mostly intended for a limited suite of applications.

An alternative approach that enables the suppression of optical diffraction is to exploit metamaterials and plasmonics [4, 13, 14]. Through leveraging the material's anisotropy, the dispersion of metamaterials can flatten in the epsilon-near-zero (ENZ) regime, which permits all wave harmonics to propagate in the same direction, allowing the fields to canalize without diffraction. On the other hand, plasmonics can also support optical fields on the subwavelength scale, facilitating the constructions of diffractionless devices in conjunction with achieving such effect with nano-scale confinement [5, 15]. This can enhance wave-matter interaction and hence passive and active integrated devices figure of merit. However, the performance of ENZ-based devices is typically compromised by the excessive material absorption that is inherent to the material constituents of such devices, which drastically curtails the propagation length of the canalized beams [16]. In contrast, metamaterials with ultrahigh material absorption are capable of supporting low-loss canalized field [17, 18]. In this case, the optical field overlapping lossy materials can be minimized, reminiscent of perfect electric conductor (PEC) effect where the electric field resident in PEC can diminish due to infinite material absorption. Nevertheless, this technique will severely impact optical field confinement, and the high material absorption is also not directly achievable in most of the natural anisotropic materials, hence further implementation of artificial metastructures is required [18-20]. Recently, by stacking two hyperbolic metasurfaces, a highly-confined, low-loss optical canalization scheme based on Moire physics has been demonstrated [21, 22]. However, the dispersion band depends sensitively on the rotation angle between the two materials, and the operation of canalization regime can only be attained within a narrow range of twisted angle.
In this work, we demonstrate theoretically that with the use of only one layer film anisotropic 2D material [23-25], one can achieve a previously unexplored design strategy capable of supporting low loss canalization of highly confined plasmon waves. By leveraging material absorption which is naturally intrinsic in these materials, it is found that in addition to elliptic and hyperbolic dispersion regimes which are common in ordinary metamaterials, this class of 2D material can possess a combination or a hybrid of elliptic and hyperbolic characteristics. This allows dispersion contours to flatten when transitioning from closed to open dispersion regimes. Notably, such a transition mechanism is not induced by the change in sign of permitivitty in ENZ material or by the control of rotation angle that dictates the hybridization of two metasrufaces, but through the strength of the material absorption. As such, material absorption, which generally represents an unwanted shortcoming for plasmonic structures, can instead be positively exploited to tailor dispersion characteristics of anisotropic 2D material structures and enable new optical functionality. Moreover, the material absorption introduced here will not negatively impact the modal loss and confinement of the plasmonic waves, and the level of material absorption required for canalization is readily available in natural 2D materials such as black phosphorus (BP) when compared to using PEC effect as shall be demonstrated. The possibilities of dispersion engineering of plasmon modes in 2D anisotropic materials thus render them a promising integrated plasmonic platform for diffractionless optics, while allowing strong tunability over device operations using gate voltage.

## Results and Discussions

**Dispersion engineering in anisotropic 2D materials**

Plasmonic waveguides, constructed using 2D plasmonic materials have emerged as a popular optical platform for the realization of photonic devices on a deep subwavelength scale. Plasmonic waveguides using 2D layers can offer the highest level of modal confinement, unprecedented device tunability and enhanced local density of states [14, 26]. In particular, 2D plasmonic materials with anisotropic optical properties have received significant attention due to their potential to connect and concomitantly profit from the fields of plasmonics and metamaterials. The versatile dispersion characteristics supported in these 2D structures, enabled by the material's anisotropy, make them an ideal platform for metamaterial-based functions and metacircuit [27] realized in a 2D flat surface. In this work, we will discuss the opportunities afforded by leveraging material absorption to achieve flat band dispersion within anisotropic 2D materials, which can allow canalization of optical fields in a 2D integrated setting. The dispersion relation of an anisotropic 2D film with in-plane anisotropic conductivity tensor $\sigma_{x,z}$, can be characterized by the following equation [28]:

$$(k_x^2 - k_0^2)\sigma_x + (k_z^2 - k_0^2)\sigma_z - 2i\sqrt{k_x^2 + k_z^2 - k_0^2}\left(\frac{1}{\eta_0} + \frac{\eta_0 \sigma_x \sigma_z}{4}\right) = 0 \qquad (1)$$

It is instructive to first consider the effect of material absorption on the dispersion properties in isotropic 2D materials such as graphene, where the optical properties are characterized by material's in-plane conductivity ($\sigma$) as schematically shown in Fig. 1(a). In the non-retarded regime, the isotropic 2D dispersion can be obtained by assuming $\sigma_x = \sigma_z = \sigma$ in Eq. 1, which yields [29]:

$$\sqrt{k_x^2/k_0^2 + k_z^2/k_0^2} = \frac{2}{\eta_0}\frac{\sigma_i + i\sigma_r}{|\sigma|^2} \qquad (2)$$

The plots in Figs. 1(b) and (c) display the effect of material absorption on isotropic dispersion based on Eq. 2. Without loss of generality, here we assume that the field is propagating in the z direction, and the dispersion behaviors of $k_z$ are plotted as function of Re[$k_x$], which represents waves of different harmonics. It can be seen that the introduction of material absorption (Re[$\sigma$]) will lead to the reduction and enhancement in Re[$k_z$] and Im[$k_z$] respectively, which negatively impacts modal confinement and propagation distance of 2D isotropic plasmons.

In contrast, the mechanism for anisotropic dispersion of 2D materials with material absorption becomes more complex, especially in the regime where Re[$\sigma_x$]>Im[$\sigma_x$]. Initially, we study the dispersion behavior when $\sigma_x$ is a real quantity (Im[$\sigma_x$]=0). As detailed in the supporting material, the dispersion relation in this case can be expressed as:

$$\sigma_z k_{zr}^4 - \frac{\sigma_x^2}{4\sigma_z}k_x^4 - \frac{2}{\eta_0}\sqrt{\frac{\sigma_x}{2\sigma_z}}|k_x|k_{zr}^2 = 0 \qquad (3)$$

It can be observed that when Im[$\sigma_x$]=0, the dispersion of anisotropic 2D materials can be characterized by a biquadratic equation, which exhibits parabolic features similar to those of hyperbolic dispersion as displayed in Fig. 1(h). It is noteworthy to point out that similar dispersion behavior can also be observed in 3D metamaterials operated in the ENZ regime ($\text{Im}[\varepsilon] > \text{Re}[\varepsilon] \sim 0$). In this case, the dispersion can become parabolic even though the real part of permitivity tensor is positive.

The evolution of dispersion of the $\text{Re}[k_z]$ with respect to material absorption ($\text{Re}[\sigma_x]$) is further displayed in Figs. 1(e)-(h). It is shown that anisotropic 2D materials can exhibit elliptic, parabolic, and a mixed type of dispersion depending on the value of $\text{Re}[\sigma_x]$. In Figs. 1(f) and (h), the dispersion contours are closed and open curves in the regimes where $\text{Re}[\sigma_x] \ll \text{Im}[\sigma_x]$ and $\text{Re}[\sigma_x] \gg \text{Im}[\sigma_x]$ respectively. On the other hand, the mixed dispersion can occur with moderate material absorption ($\text{Re}[\sigma_x] \approx \text{Im}[\sigma_x]$), which exhibits a combination of elliptic and parabolic dispersion as displayed in Fig. 1(g).

As evidenced from the results, material absorption offers a path toward dispersion engineering, making it possible to transition from close to open dispersion through the leverage of the strength of material absorption. Such a transition can be verified in Figs. 1(i)-(k), which illustrates the field distributions in 2D anisotropic material launched by a dipole polarized in the out of plane direction. It is shown that a more diffracting, elliptic wavefront is observed in the low material absorption regime, whereas a more confined, hyperbolic-like wavefront can be observed in the high material absorption regime. In the intermediate regime, the anisotropic 2D materials instead can support an elliptic, but a more confined wavefront as depicted in Fig. 1(j); a feature corresponding to mixed type of dispersion. It should be highlighted that such a feature cannot be obtained in isotropic materials, where the wavefront is always circular in nature regardless of the strength of material absorption as shown in Fig. 1(d).

It is noteworthy to highlight two important attributes of the mixed type dispersion supported in such anisotropic 2D materials. Because the contour depicts elliptic-like dispersion for smaller $k_x$ and parabolic-like dispersion for larger $k_x$, there exists a transition point at which the slope ($dk_z/dk_x$) of the curve becomes zero. As detailed in the supporting materials, the value of $k_x$ corresponding to this point can be obtained by assuming $dk_z/dk_x = 0$ in Eq. 1, which is given by:

$$\frac{k_x}{k_0} = \frac{1}{\eta_0 |\sigma_x|} \sqrt{\frac{Im[\sigma_x]}{Im[\sigma_z]}}. \qquad (4)$$

On the other hand, in the high $k_x$, non-retarded regime, the slope of dispersion curve in the parabolic/hyperbolic-like regime is also derived in the supporting material, and can be expressed as:

$$\frac{dRe[k_z]}{dk_x} = \sqrt{\frac{Re[\sigma_x]}{2Im[\sigma_z]}}. \qquad (5)$$

From Eqs. 4 and 5, it is seen that in this case the slope of the curve in the high $k_x$ regime is predominantly characterized by material absorption, which in turn can confirm the dispersion curve obtained in the high $\text{Re}[\sigma_x]$ regime in Fig. 1(h).

In the field of metamaterials, similar mixed/hybrid dispersion can also occur in bianisotropic materials [30], which are described by high order dispersion relations with circularly or elliptically polarized eigenwaves, or parity time (PT) symmetry metamaterials [31] which involve spatial modulation of loss and gain. However, these approaches involve the interaction of light with subwavelength nanoparticles and are mostly intended for free space applications due to the flow of light from and into surfaces orthogonally, rather than in-plane. In contrast, naturally occurring 2D materials are better suited as an integrated in-plane platform due to their structural simplicity. The mixed type of dispersion afforded in anisotropic 2D materials therefore can facilitate new integrated optical functionalities that are not possible in conventional elliptic and hyperbolic metamaterials. Markedly, it indicates the possibility of topological transition from elliptic to hyperbolic dispersion as a functional regime where devices can be built and operated. Such a transition is of great interest to plasmonic metadevices because the dispersion contour can be tailored by regulating material absorption. In next section, we will demonstrate how to further leverage such dispersion characteristics to achieve plasmon canalization in 2D materials for diffractionless optics.

**Field canalization in anisotropic 2D materials facilitated by material absorption.**

The mixed type dispersion in anisotropic 2D materials can provide new opportunities to achieve optical functions unattainable in their isotropic counterparts. In particular, the existence of a transition point along with the control over the slope of the curve in the open regime, allow the dispersion to flatten by leveraging material absorption. Because the slope in the elliptic and hyperbolic regimes are predominantly dictated by degree of anisotropy ($Im[\sigma_x]/Im[\sigma_z]$) and the material absorption respectively [Eq. 5], one can define the following condition to achieve flat band dispersion in anisotropic 2D materials:

$$Im[\sigma_z] > Im[\sigma_x] \approx Re[\sigma_x]. \qquad (6)$$

Figures 2(a)-(b) illustrate an example as to how to leverage material absorption to achieve flatband dispersion based on Eq. 6. It is seen that for an anisotropic 2D film free of material absorption ($Re[\sigma_x] = 0$), the dispersion curve will exhibit an elliptic characteristics when $Im[\sigma_{x,z}]$=0.1,0.2imS as depicted by the blue curve in Fig. 2(a). However, by increasing the strength of material absorption such that $Re[\sigma_x] \approx Im[\sigma_x]$, the initial, closed, elliptic dispersion contour can become open and the mixed type of dispersion will be supported. The opening angle of the curve can be tuned by using various values of $Re[\sigma_x]/Im[\sigma_z]$ as depicted by the red ($\sigma_x$=0.1i+0.06mS) and black ($\sigma_x$=0.4i+0.3mS) curves. On the other hand, the dispersion curve will become predominantly hyperbolic when $Re[\sigma_x]$ further increases, losing the elliptic feature as plotted by the green curve in Fig. 2(a)

Such a material absorption-induced transition can be further examined in Fig. 2(b), which details the dispersion characteristics in the regime wherein $-100 < k_x/k_0 < 100$. Due to the emergence of a transition point pertaining to the mixed type of dispersion, a more flattened contour curve can be obtained in this regime when $Im[\sigma_x] < Im[\sigma_z]$ with the use of moderate material absorption as given by Eq. 5. This dispersion feature, in turn, can facilitate field canalization in anisotropic 2D materials, which requires the Poynting vector of different $k_x$ modes to have the same direction. As depicted in Fig. 2(c), the Poynting vector can become unidirectional ($\theta = 0$) for a broad range of $k_x$ modes when $Re[\sigma_x] \approx Im[\sigma_x]$, hence allowing modes of different spatial harmonics to canalize in the same propagation direction.

When launched by a dipole source, the canalization effect enabled by flattened dispersion can suppress the diffraction of the propagating field, permitting a broad range of $k_x$ harmonics to propagate in the same direction. To assess the effect of canalization induced by material absorption, we compare the canalization angle, defined by ΔFWHM/Δd [22], as function of $\text{Re}[\sigma_x]$ as plotted in Fig. 2(d), where FWHM is the full width half maximum of the wavefront and Δd is the propagation distance of the fields. The expansion of the wavefront can be optimally suppressed when $\text{Re}[\sigma_x] \approx 0.6\text{Im}[\sigma_x]$. In this case, the canalization angle can be as small as 13º, indicative of the possibility of field canalization through the regulation of material absorption.

It is important to note that diffraction can be further suppressed through the interplay between material absorption and anisotropy for the best performance of canalization. As displayed in Fig. 2(f), it is seen that with the same level of material absorption ($\text{Re}[\sigma_x] = 0.6\text{Im}[\sigma_x]$) being used, the canalization angle can be reduced further by increasing the degree of material anisotropy ($\text{Im}[\sigma_z]/\text{Im}[\sigma_x]$). This is due to a broader range of canalized $k_x$ modes achievable in anisotropic 2D materials when using a stronger degree of material's anisotropy on top of material absorption, as illustrated in Fig. 2(e). Also shown in the plot is the open angle, which can offer an analytical estimation of the canalization effect based on the slope in the hyperbolic regime ($\tan^{-1}(\text{Re}[\sigma_x]/2\text{Im}[\sigma_z])$). It is evident that the trend in the canalization angle shows good agreement with that of open angle, which suggests that the stronger canalization effect is because dispersion can be flattened into the higher $k_x$ regime as shown by the green curve in Fig. 2(e).

**Loss and confinement tradeoff in 2D canalized structures**
In this section, we will show how the material absorption-induced flatband dispersion can be engineered to enable low loss, highly confined canalized fields when compared to other canalized techniques based on ENZ and PEC effects. This can address the design bottleneck that is common in plasmonic devices, where material absorption typically dictates strict trade-offs between modal loss and confinement. For instance, for canalization through using ENZ effects, the material absorption will greatly reduce the propagation length of the canalized plasmon fields whereas in the PEC scheme, low loss canalized plasmon fields can be obtained but subwavelength confinement of the modes will be compromised. It should be noted that similar design trade-off has been recently addressed using composite hybrid plasmonic structures, which can facilitate device performance of various plasmonic functional devices [32, 33]. Anisotropic 2D materials can instead offer a platform to achieve superior canalized mode attributes without the need for implementing layered waveguide structures, while simultaneously offering strong device tunability.

Counter-intuitively, the low loss canalized field achieved in this work is facilitated by regulating material absorption such that $\text{Re}[\sigma_x] \approx \text{Im}[\sigma_x]$. This is in stark contrast to conventional design wisdom, which dictates that the formation of low loss plasmonic modes should require the functional plasmonic materials be operated either in low or ultrahigh material absorption regimes, such as the ones in ENZ and PEC canalized schemes respectively. To understand how moderate material absorption can impart such a low loss canalization effect, in the supporting material we study the dispersion behavior of the real and imaginary components of $k_z$ as a function of $k_x$. As can be seen, the material absorption introduced here has a disproportional effect on the modal loss in the elliptic and parabolic regimes, as $\text{Im}[k_z]$ increases more significantly only after the transition point into the hyperbolic regime. As such, a broad range of low loss $k_x$ modes can still be achieved in the elliptic regime, which in turn permits low loss canalized field propagation.

An important figure of merit for plasmonic modes is the ratio of $\text{Im}[k_z]$ over $\text{Re}[k_z]$ [15,34], which reflects the length of propagation distance of the plasmon modes normalized to the wavelength of operation. Additionally, confinement of the modes can be estimated by the penetration depth (*L*) into the vacuum, as defined by $\text{Im}[k_y]/2\lambda_0$. Such a metric has been utilized widely to characterize the confinement of plasmonic modes. Based on these two metrics, we can compare the modal loss and confinement of different 2D canalized schemes as functions of material absorption and $k_x$, which are plotted in Fig. 3.

As displayed in Figs. 3 (a)-(c), the increased material absorption in the scheme we describe in this work can enable a relatively broadband, low loss dispersion in the elliptic operating regime. Furthermore, the modal confinement is nearly unaffected, indicating that all spatial harmonics can still carry deeply-subwavelength information, despite the presence of material absorption.

Conversely, the material absorption in ENZ canalized scheme represents an undesired shortcoming to the quality of plasmon modes. As can be seen in Figs. 3 (d)-(f), the increased material absorption will negatively impact the modal loss and confinement, therefore degrading the performance of canalized plasmon fields. As such, material absorption should be minimized for the best performance of ENZ canalization. On the other hand, in the PEC scenario shown in Figs. 3(g)-(i), the 2D materials should be operated in the ultrahigh regime to support low loss plasmon canalized fields. However, as displayed in Fig. 3(i), the modal confinement will be severely compromised in the low $k_x$ regime, inevitably losing subwavelength features originally offered by 2D plasmoncis. Similar design trade-off associated with ENZ and PEC canalized schemes can also be observed in their 3D metamaterial version as discussed in the supporting material. Overall, as showcased in Figs. 3(a), (d) and (g), the afforded mixed type of dispersion can better alleviate the associated loss-confinement trade-off by moderately regulating material absorption, offering a new avenue for the construction of diffractionless devices with improved performance. Compared with other canalized platforms, the material absorption inherent in anisotropic 2D materials can therefore favor low loss plasmon canalization in the deep subwavelength regime.

**Anisotropic 2D materials as a gate-tunable canalization platform**

In this section, we will demonstrate how to leverage material absorption of natural 2D anisotropic materials such as BP to realize low loss, highly confined canalized modes. Recently, BP plasmonics operating in the mid-infrared (MIR) regime have received significant interest for nanophotonic applications such as sensing, molecular analysis and optical communications [23, 35]. Here we shall showcase that it can also serve as a reconfigurable, low loss canalized platform in these important wavelength regimes. One prominent advantageous feature of anisotropic 2D material over bulk metamaterials is the ability to tune the conductivity tensor by chemical potential or electric bias, which can offer strong tunability over wave front manipulation in BP film as schematically depicted in Fig. 4(a). For instance, Fig. 4(b) displays the conductivity tensor of 20nm BP as a function of chemical potential in the MIR regime when $\lambda_0 = 3.25\mu m$, where x and z are the armchair and zigzag directions of BP respectively. The strength of material absorption in $\sigma_x$ can be tuned significantly as a function of applied chemical potential. The canalization condition based on Eq. 6 can be therefore achieved in the vicinity of $\mu = 0.08\text{eV}$, where $\text{Re}[\sigma_x] \approx \text{Im}[\sigma_x]$.

A dynamical topological transition from elliptic ($\mu = 0.1eV$) to hyperbolic ($\mu = 0.06eV$) contours can be realized, while the transition point corresponds to flattened dispersion ($\mu = 0.08eV$) as shown in Fig. 4(c). BP film operated with this condition can enable highly-collimated canalized fields as illustrated in the field distributions within Fig. 4(e). In this case, BP can support long range, highly canalized field with a diffraction angle of 5.5° and propagation distance of $5\lambda_{spp}$, highlighting BP's potential as a tunable diffractionless material platform. It should be noted that the transition mechanism here is induced by regulating the material absorption, provided that the dispersion of the functional anisotropic materials can sweep across the point where the strength of material absorption is moderate ($\text{Re}[\sigma_x] \approx \text{Im}[\sigma_x]$). On the other hand, the diffraction angle of BP operated under the ENZ condition ($\mu = 0.075eV$) will increase to 15° with a reduced propagation distance of $2\lambda_{spp}$. In this case, the dispersion curve will become hyperbolic-like which will lead to a more diffracting field propagation. Since the strength of material absorption can be tuned through chemical potential or bias in 2D materials, such a transition can be viable in most of the anisotropic 2D materials but not limited to BP.

Table. 1 further compares the optical performance of various canalization platforms, where metamaterials can be realized using metal-dielectric structures [19, 20]. There are at least two major dispersion characteristics that are distinct from those of BP counterpart. First, the range of $k_x$ modes in the canalization regime in metamaterials ($-5 < k_x/k_x < 5$) is about two order magnitude smaller than that of BP ($-300 < k_x/k_x < 300$), which hinders higher $k_x$ harmonics from contributing canalization process. As such, BP plasmonics allow for better near-diffractionless propagation as compared with metamaterials, with diffraction angle of only 5°. Second, the value of $Re[k_z]$ in the canalized regime in metamaterials based on metal is also around two order magnitude smaller than that of BP. The capability of 2D materials such as BP to support highly confined surface mode with guided wavelength ($\lambda_{spp}$) much smaller than $\lambda_0$ make them a more favorable platform of choice for one to realize canalized fields in strongly subwavelength regime. In contrast, the guided wavelength supported in metamaterial or metal plasmonics will be within the same order as $\lambda_0$. In order to, account such subwavelength feature enabled by 2D materials, the figure of merit for the modal loss of 2D plasmonic is defined by $Im[k_z]/Re[k_z]$, which indicates the propagation distance normalized to guided wavelength [15]. Clearly, through the regulation of material absorption, anisotropic 2D plasmonic material can represent a new paradigm for the constructions of diffractionless devices, while simultaneously facilitating long range, highly canalized field in the deep subwavelength regime.

**In-plane, integrated canalization platform in 2D material structures**

In this section, we will discuss the attributes of anisotropic 2D materials as an attractive integrated photonic platform where field canalization can be configured in an in-plane, guided wave setting without the need for combining other cladding layers to create 3D stacks. Integrated photonic devices can offer greater stability, density and scalability than their free space counterpart. Most importantly, by exploiting the anisotropic optical properties, these anisotropic materials can play the role of optical nanocircuit elements analogous to microelectronics [27]. Thus, an integrated platform that can support low loss, subwavelength field canalization with dynamic reconfigurability is highly desired. However, these important attributes may not be obtainable simultaneously with the use of current technological platforms.

For instance, with the use of twisted bilayer hyperbolic metasurfaces such as α-MoO$_3$, a tunable, low loss, deeply subwavelength canalized plasmon scheme can be achieved [21]. Such a platform is based on the dispersion hybridization between anisotropic plasmon polartion fields in stacked α-MoO$_3$, reminiscent of the peculiar electronic band structure modified by van-der-Waals superlattices [36]. More recently, it is even demonstrated that such a bilayer twisted structure can be deployed to effectively create longitudinal spin of plasmon necessary for chiral plasmonics [37, 38]. However, despite the numerous emerging photonic applications that twisted optics can enable, the requirement of stacking two films in the out-of-plane direction makes it more intended for free space elements. Additionally, the dispersion of such bilayer structures depends sensitively on the twisted angle between the films, which is not tunable therefore limits its capabilities for active applications. On the other hand, a coplanar scheme based on PEC [4] effect has also been proposed to achieve plasmon canalization. However, the operation of this scheme should require the implementation of composite subwavelength metastructures to achieve the desired material parameters. The actual canalization performance, based on such periodic structures, will be restricted by optical nonlocality [25], particularly in the in-plane, integrated setting where such nonlocal optical effect will be more pronounced with the presence of in-plane wave vector. Additionally, similar to bilayer hyperbolic canalized scheme, the implementations of densely packed multilayer structures in these platforms will lead to higher optical loss more stringent to material granularity and impurity, which degrades the quality of canalized plasmon fields. In contrast, the canalization scheme facilitated by material absorption proposed in this work will not have these challenges as flatband engineering can be accomplished with the use of only single anisotropic 2D film which can be gate tunable in an in-plane, integrated setting.

To highlight the possibility of anisotropic 2D material as an integrated canalization platform, an in-plane excitation scheme involving an unbounded BP film connected to a BP ribbon waveguide of finite width can be considered, as schematically depicted in Fig. 4(g). BP can suppress optical diffraction even when the ribbon mode (z<0) enters the unbounded regime (z>0), maintaining highly confined Gaussian field profiles as the mode propagates down further into the film when $\mu = 0.08eV$. It should be noted that although similar diffractionless field propagation has also been reported using graphene solitons enhanced by graphene plasmonics [39], no nonlinear effect is required in this work as the canalization is due to flatband dispersion enabled by material absorption.

Compared to other integrated diffractionless techniques, the flatband engineering in 2D anisotropic materials can allow field canalization in 2D integrated plasmonic without the need for combining other 3D structures. For instance, propagation of diffraction-free Bessel beam in an integrated, guided-wave setting has been recently reported [10]. However, the generation of such Bessel-type beam based on axicon lens approach requires delicate controls over the phase and amplitude of the propagating waves. As such, phase controlling elements such as optical nanoresonators or photonic crystals should be deployed [10, 40] for the operation of axicon lens. The inclusions of these resonator structures will lead to higher losses, bulkier device footprint, and restricted bandwidth of operation with limited device reconfigurability. Contrarily, our platform based on flatband engineering can represent a more advantageous platform as no such supporting structures is needed in the design. As a whole, anisotropic 2D materials can offer a planar 2D integrated canalization scheme that is readily available in naturally occurring materials, while simultaneously facilitate low loss and subwavelength plasmon propagation with no need to relying on additional 3D structures.

In summary, we have elucidated a physical mechanism where material absorption can be exploited to facilitate dispersion engineering in anisotropic 2D materials. It is found that by regulating material absorption, the dispersion contour can flatten, allowing low loss plasmon canalization in the subwavelength regime with strong tunability. Some of these capabilities can be achieved in other 2D or 3D metastructures, but not with the dynamic reconfigurability nor performance that a single layer of anisotropic 2D material provides. The realization of such canalization effect can be immediately considered using natural anisotropic 2D materials such as BP. Because the conductivity and the strength of material absorption of BP can be dynamically configured through bias or chemical potential, topological transition process therefore can be observed via bias using BP film. When the dispersion flattens, BP can support low-loss canalization field with diffraction angle of $5.5^{\circ}$. In addition to near-diffractionless propagation, we also envision that the versatility of dispersion supported in this class of 2D materials, namely capable of supporting elliptic, hyperbolic-like, and mixed type of contour behaviors through bias or doping, can represent a promising material platform to achieve new functionalities for 2D plasmonic metadevices.

## Methods

### Optical modelling and simulations of 2D plasmonics

The dispersion plots are obtained by solving Eq. [1] using *Mathematica.* The simulations of the dipole radiation profile are conducted using *Lumerical FDTD*, where the electric dipole is placed at the center of the film. The mesh size was set to be $\lambda_0/5000$ around the film due to the deeply subwavelength feature of 2D materials.

40. Zhu, L. et al. "Electrically-pumped, broad-area, single-mode photonic crystal lasers," *Opt. Express* **15**, 5966–5975 (2007).

**Figures and Tables**

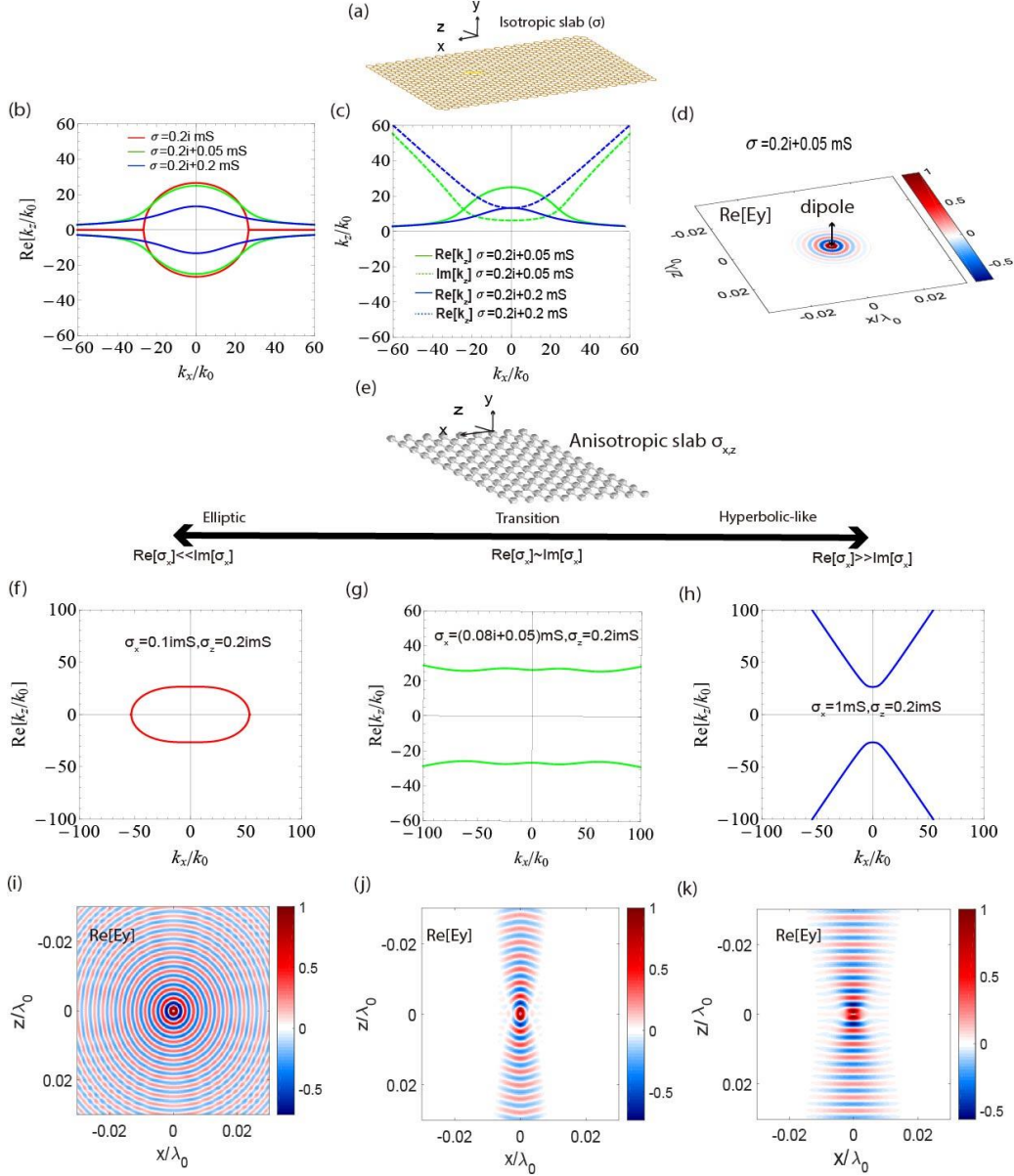

**Fig. 1. Dispersion of anisotropic 2D materials.** (a) Schematic of an isotropic 2D material sitting on the x-z plane, which is optically characterized by in-plane conductivity $\sigma$. Dispersion contours of (b) $\text{Re}[k_z]$ and (c) $\text{Im}[k_z]$ of an isotropic 2D material as function of $k_x$ using various values of $\text{Re}[\sigma]$. It is seen that the increase in material absorption ($\text{Re}[\sigma]$.) will lead to the reduction in $\text{Re}[k_z]$ and the increase in $\text{Im}[k_z]$, which is considered as a shortcoming for 2D plasmon due to reduced propagation distance. (d). For isotropic 2D films, the wave front launched by a y-polarized

electric dipole is always circular regardless of the strength of material absorption. (e). Schematic of an anisotropic 2D material characterized by an in-plane anisotropic conductivity tensor ($\sigma_{x,z}$). Contrary to isotropic 2D films, anisotropic 2D materials can support (f) elliptic, (h) hyperbolic, and (g) mixed type of dispersion depending on the strength of material absorption ($\text{Re}[\sigma_x]/\text{Im}[\sigma_x]$). For mixed dispersion, the dispersion contour can possess a hybrid character of the elliptic and hyperbolic dispersions. When launched by a y-polarized dipole, anisotropic 2D films therefore can support (i) elliptic, or (k) hyperbolic field distributions in the low and high material absorption regimes. (j). In the intermediate regime, mixed type of field distribution can be supported, which offers a pathway toward dispersion transition.

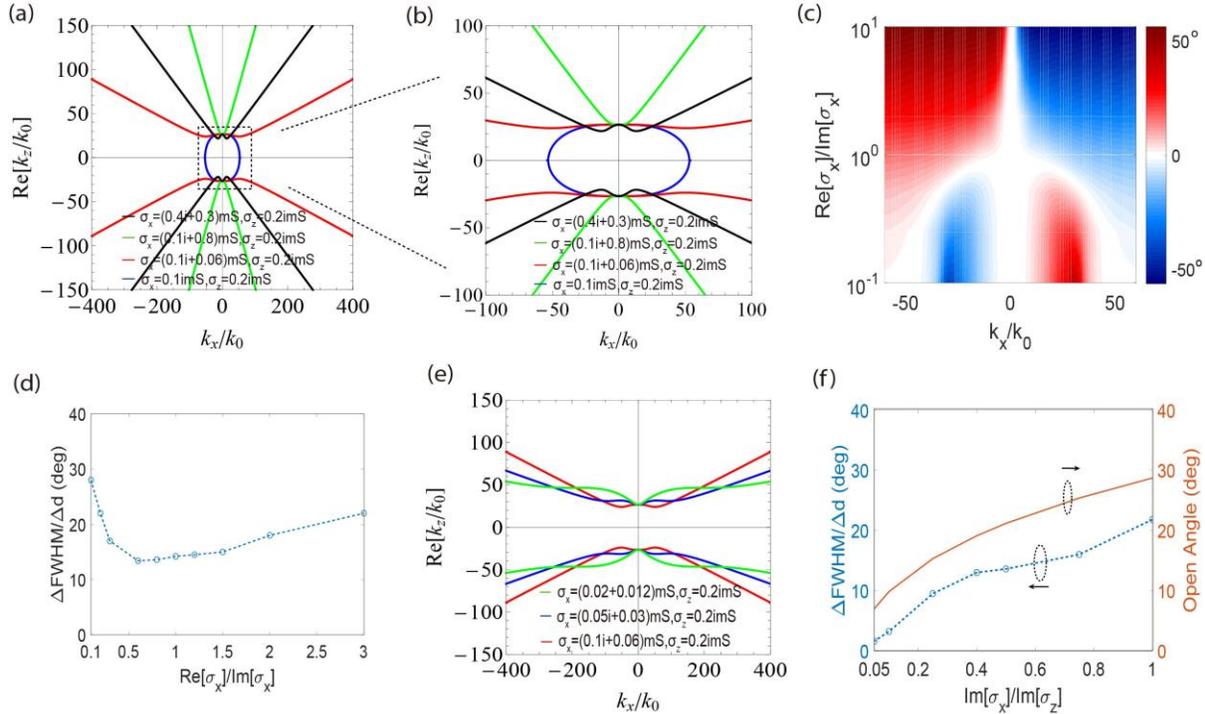

**Fig. 2. Flatband dispersion in anisotropic 2D materials.** (a) Dispersion curves of anisotropic 2D materials using various combinations of $\text{Re}[\sigma_x]$, and $\text{Im}[\sigma_x]$, which can support elliptic, hyperbolic and mixed type of dispersion by regulating material absorption, allowing contour to transition from close to open. (b) Detailed view of the dispersion curves shown in (a). It is seen that when $\text{Re}[\sigma_x]\sim\text{Im}[\sigma_x]$, the dispersion curve can be flattened in the regime where $-100 < k_x/k_0 < 100$ as a result of mixed type of dispersion. (c) The angle of the direction of Poynting vector ($\partial k_z/\partial k_x$) of $k_x$ modes as function of material absorption with respect to z axis. When $\text{Re}[\sigma_x]\sim\text{Im}[\sigma_x]$, the angle of Poynting vector can become zero over a broad range of $k_x$, permitting canalization to occur. (d) Numerically simulated diffraction angle [21] of the field propagation in anisotropic 2D materials ($\text{Im}[\sigma_{x,z}]$=0.1,0.2imS) as function of material absorption ($\text{Re}[\sigma_x]/\text{Im}[\sigma_x]$) launched by a dipole source. The dipole source is pointing in the out-of-plane (y) direction and will excite all $k_x$ spatial harmonics. It is seen that the diffraction can be best suppressed when $\text{Re}[\sigma_x] \approx 0.6\text{Im}[\sigma_x]$. (e) Dispersion curves in anisotropic 2D materials with varying degree of in-plane anisotropy ($\text{Im}[\sigma_z]/\text{Im}[\sigma_x]$) while using the same strength of material absorption ($\text{Re}[\sigma_x] = 0.6\text{Im}[\sigma_x]$). It is shown that on top of material absorption, it is possible to further flatten dispersion curve with the aide of more extreme material's anisotropy. (f) Diffraction and open angels of the propagating fields in anisotropic 2D materials ($\text{Im}[\sigma_x] = 0.2mS$ and $\text{Re}[\sigma_x] = 0.6\text{Im}[\sigma_x]$)) as function of in-plane anisotropy launched by a y-directed dipole source. Diffraction can be further suppressed by decreasing the value of $\text{Im}[\sigma_x]/\text{Im}[\sigma_z]$ as a result of more flatted dispersion enabled by anisotropy

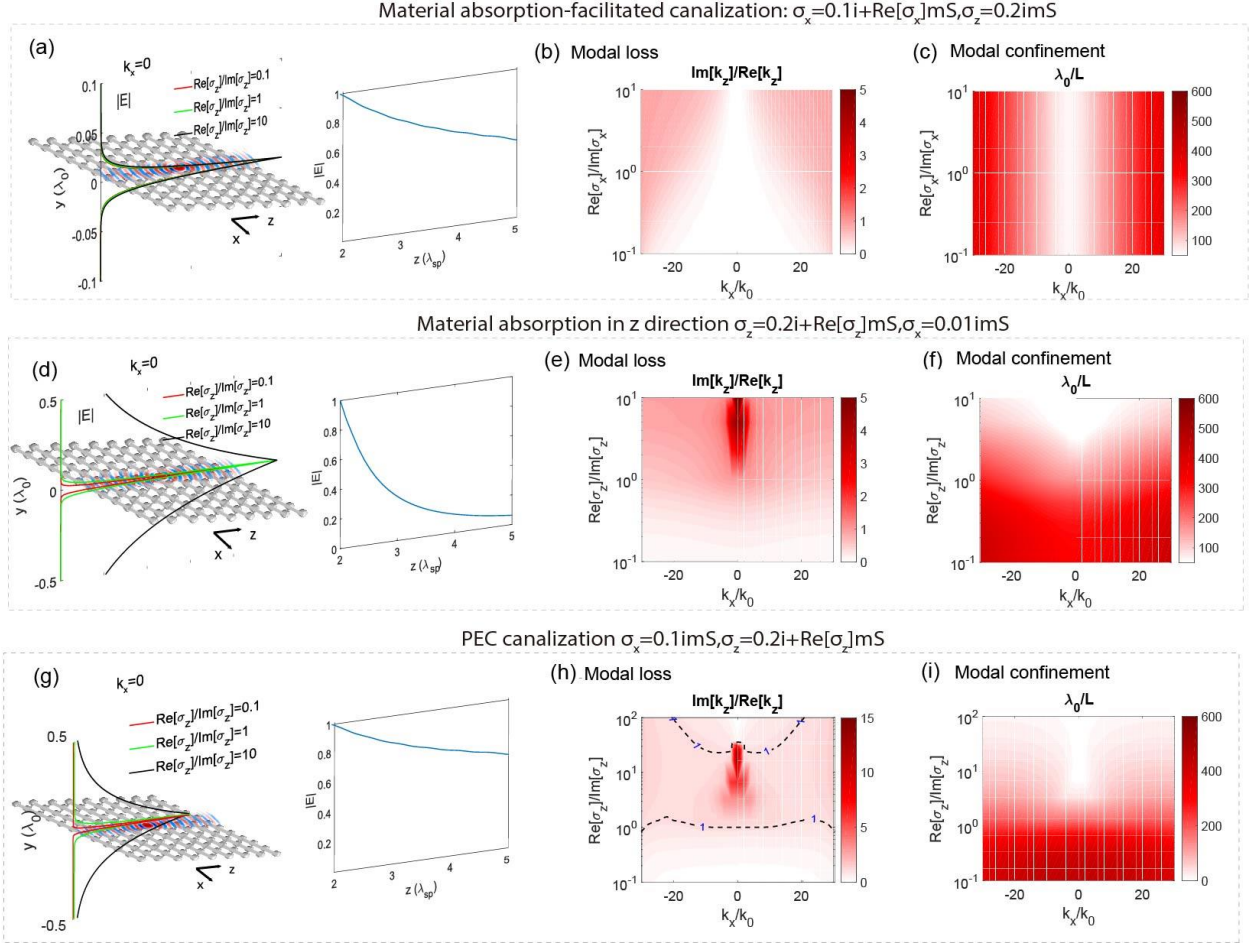

**Fig. 3. Loss-confinement for different canalized schemes.** Comparison of modal loss ($\text{Im}[k_z]/\text{Re}[k_z]$) and confinement ($L/\lambda_0$) in different 2D canalized platforms based on: (a)-(c) material absorption shown in this work (d)-(f) ENZ effect, and (g)-(i) PEC effect. Larger penetration depth (*L*) will lead to weaker modal confinement. (a) Schematic, penetration depth as function of material absorption ($k_x = 0$), and the power attenuation ($\sigma_{x,z}$=0.1i+0.06, 0.2imS) of the canalized field in anisotropic 2D materials facilitated by material absorption. (b) Modal loss and (c) modal confinement of anisotropic 2D modes as functions of $k_x/k_0$ and material absorption. It is seen that a broad range of low loss $k_x$ modes can still be supported even with the introduction of material absorption, while the modal confinement remains largely non-impacted as shown in (a). (d) Schematic, penetration depth, and the power attenuation of the canalized field facilitated by ENZ metasurface ($\sigma_{x,z}$=0.2i+0.06, 0.001imS), where it is seen that the propagation distance will be greatly curtailed with the introduction of material absorption. (e) Modal loss and (f) modal confinement of ENZ metasurface as functions of $k_x/k_0$ and material

absorption. The performance of modal loss and modal confinement will be degraded with the increase of material absorption, particularly in small $k_x$ regime. (g) Schematic, penetration depth, and the power attenuation of the canalized field facilitated by PEC effect. It is seen that the modes operated in PEC condition will become loosely guided, losing subwavelength feature offered by 2D plasmonics. (h) Modal loss and (i) modal confinement of PEC metasurface as functions of $k_x/k_0$ and material absorption. As can be seen, the modal loss can be reduced however the modal confinement will be degraded in the PEC regime.

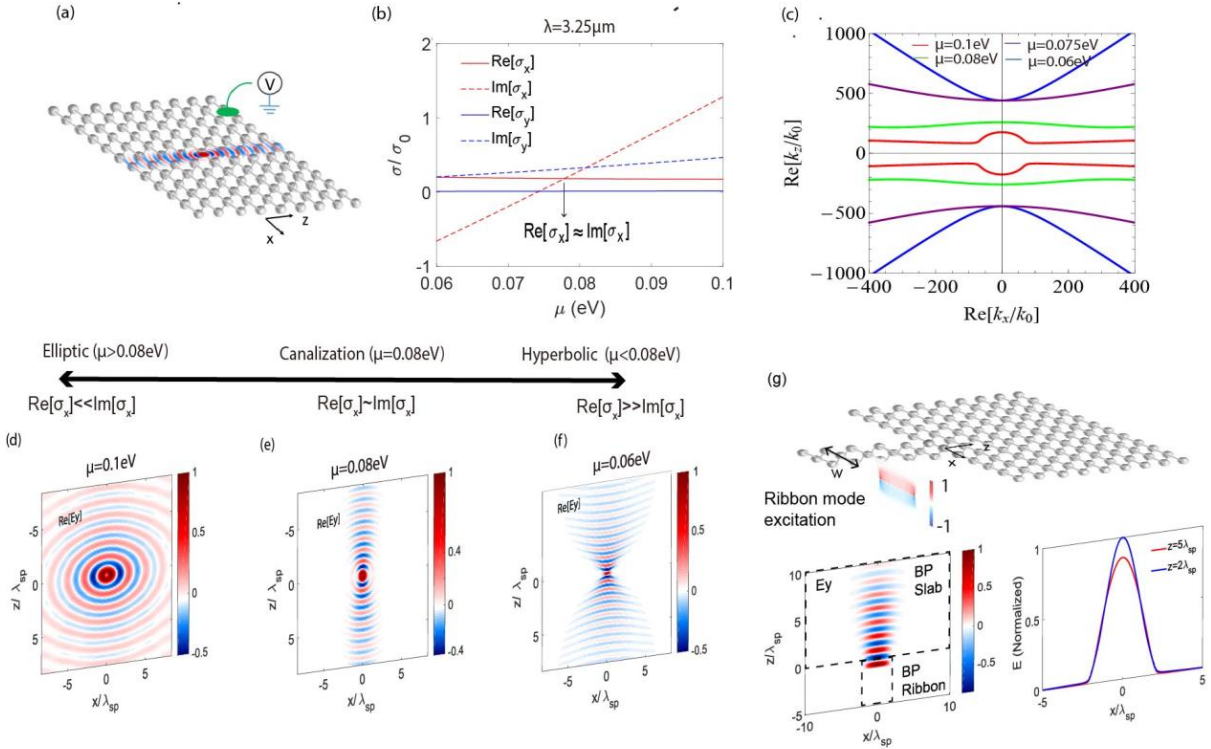

**Fig. 4. BP as a gate tunable 2D canalization platform.** (a) Schematic of BP film being considered for the realization of low loss canalization field tunable by bias or chemical potential. (b) Normalized conductivity tensor ($\sigma_0 = e^2/4\hbar$) in the armchair (x) and zigzag (z) directions for 20nm BP as function of chemical potential [23,24] when $\lambda_0 = 3.25 \mu m$. The conductivity of BP can be highly tunable where various material absorption level can be achieved via bias or chemical doping. In this case, moderate material absorption can be attained in BP when μ= $0.08eV$. (c) Material absorption-facilitated topological transition in BP when $\lambda_0 = 3.25 \mu m$. The dispersion contour can be elliptic, flattened, and hyperbolic whenμ= $0.1, 0.08, 0.06eV$ respectively, showing strong tunability over dispersion characteristics. The dispersion of BP under ENZ condition (μ= $0.075eV$) is also plotted for comparison. (d)-(f) Field distributions in such functional BP film using different chemical potential, allowing dynamical control over the dispersion and wavefronts manipulation with different diffraction characteristics. When μ= $0.08eV$, the diffraction angle is calculated to be only 5.5º, which can support near-diffractionless propagation of plasmon field. (g) Demonstration of an in-plane, integrated canalization platform using BP (μ= $0.08eV$) when excited by a ribbon mode of finite width, where the Gaussian field profile of the ribbon mode supported in BP ribbon waveguide (z<0) can be

Manuscript Template                                           Page **16** of **17**

maintained even in the regime when the width of BP becomes unbounded (z>0), exhibiting diffractionless feature in an in-plane setting.

| Canalized Scheme | Platform | Diffraction angle | Propagation length | Guided wavelength ($\lambda_{spp}$) | Penetration depth |
|---|---|---|---|---|---|
| Metamaterial (PEC) [20] | Free space | 16º | $2.5\lambda_{spp}$ | $\approx \lambda_0/3$ | $\approx \lambda_0/40$ |
| Metamaterial ENZ | Free space | 10º | $0.25\lambda_{spp}$ | $\approx \lambda_0/1.2$ | $\approx \lambda_0/15$ |
| BP (PEC) [18] | Integrated | 10º | $0.75\lambda_{spp}$ | $\approx\lambda_0$ | $\approx \lambda_0/10$ |
| Bilayer metasurfaces [22] | Free space | 6.6º | $5\lambda_{spp}$ | $\approx\lambda_0/50$ | $\approx \lambda_0/500$ |
| BP (Material absorption) | Integrated | 5.5º | $5\lambda_{spp}$ | $\approx\lambda_0/300$ | $\approx \lambda_0/4000$ |

**Table 1. Performance comparison of various canalization platforms.** The construction of metamaterial based on PEC effect is realized using layered structures involving zirconium nitride (ZrN) and a dielectric with $\epsilon = 10$. [20], while the metamaterial based on ENZ effect can be considered involving AZO and TiO2. PEC scheme based on BP can also be considered with the use of periodic BP nanoribbons [18]. Unlike BP plasmonics, these platforms are unable to support modes in the deep subwavelength regime. Penetration depth of the modes ($k_x = 0$) is also shown.